\documentclass[prb,twocolumn,amsmath,amssymb,superscriptaddress]{revtex4}
\usepackage{graphicx}
\usepackage{bm}
\usepackage{amsmath}

\input{tcilatex}

\begin{document}

\title{Spin relaxation in quantum dots with random spin-orbit coupling}

\author{E. Ya. Sherman }

\affiliation{Department of Physics and Institute for Optical Sciences,
University of Toronto, 60 St. George  St., Toronto M5S 1A7, Ontario, Canada}

\author{D. J. Lockwood}

\affiliation{Institute for Microstructural Sciences, National Research Counsil of Canada, 
1200 Montreal Road, Ottawa, ON, K1A 0R6, Canada}

\begin{abstract}
We investigate the longitudinal spin relaxation arising due
to spin-flip transitions accompanied by phonon emission in quantum dots where the
strength of the Rashba spin-orbit coupling is a random function of the lateral
(in-plane) coordinate on the spatial nanoscale.
In this case the Rashba contribution to the
spin-orbit coupling cannot be
completely removed by applying a uniform external bias across the quantum dot
plane. Due to the remnant random contribution, the spin relaxation rate
cannot be decreased by more than two orders of magnitude even when the
external bias fully compensates the regular part of the spin-orbit coupling.
\end{abstract}


\maketitle

\section{\protect\smallskip Introduction}

A quantum degree of freedom of an electron localized in a quantum dot (QD), i.e. its spin, is thought 
to be a useful tool for the realization of nanoscale devices that can be used for information processing.\cite
{Burkard99,Elzerman04} The interaction 
of spins with the environment  on the one hand allows necessary {\it read and write} procedures on the
other hand, leads to losses of the information held by the system. 
Thus, a critical issue regarding the possibility to convert quantum dots 
into a hardware realization of quantum information devices is the 
the ability to manipulate quickly the spins of electrons localized in
quantum dots and to keep them in the desired states as long as
necessary.  The spin-orbit (SO) coupling in QDs plays a crucial
role both for the spin manipulation and lifetime of the prepared spin
states. For example, the SO coupling allows the effective
manipulation of spins by an external electric field due to the fact that the
electric field influences the orbital degrees of freedom, and, through the
SO coupling, the spin states. The most interesting example of such a
manipulation is the electric dipole spin resonance \cite{Rashba62,Rashba03a}, the effect that occurs when the electric field of the incident
electromagnetic wave causes spin-flip transitions resonating with the wave
frequency. In this case the electric field is a much more efficient tool for
manipulating the spins than the magnetic field. The spin states in quantum
dots can be prepared and controlled by an external optical field \cite
{Sham04} too, thus allowing an optical realization of the $write\ $and $read$
operations for applications in information technologies.

The SO coupling in two-dimensional systems based on (001)-type 
structures  is described by the sum
of the Rashba \cite{Rashba84,Rashba60} 
$\hat{H}_{\mathrm{R}}=\alpha _{\mathrm{R}}\left( \sigma _{x}\hat{k}_{y}-\sigma _{y}\hat{k}_{x}\right)$ and
Dresselhaus-originated\cite{Dyakonov86} 
$\hat{H}_{\mathrm{D}}^{(001)}=\alpha_{\mathrm{D}}^{(001)}\left(\sigma_{x}\hat{k}_{x}-\sigma_{y}\hat{k}_{y}\right)$ 
terms, where $\alpha _{\mathrm{R}}$ and $\alpha_{\mathrm{D}}^{(001)}$ are the
coupling constants, ${\sigma}$ are the Pauli matrices, and $\mathbf{k}%
_{\Vert }=-i\nabla_{\parallel}-(e/c\hbar)\mathbf{A}_{\parallel}$ is the
in-plane momentum of electron. Here $e$ is the electron charge, and 
$\mathbf{A}$ is the vector-potential of the external field. $\hat{H}_{\mathrm{R}}$
and $\hat{H}_{\mathrm{D}}$ terms arise due to the artificial macroscopic
asymmetry of the structures and due to the microscopic inversion asymmetry
of the unit cells, respectively. For holes the SO Hamiltonian is more
sophisticated leading to a more complicated spectra of spin excitations and
spin dynamics \cite{Rashba88,Mauritz99,Winkler00,Schliemann05,Bernevig05}.

In GaAs/Al$_{x}$Ga$_{1-x}$As structures \cite{Stein83,Jusserand95} and
Si-based transistors\cite{Wieck84}, the SO coupling constants typically
range from 10$^{-10}$ to 10$^{-9}$ eV$\cdot $cm. It is important to mention
that by applying an external bias across the quantum well, it is possible to
manipulate the magnitude of $\alpha $ in InGaAs/InAlAs-based \cite{Nitta97}
and GaAs/AlAs-based \cite{Knap96,Miller03,Karimov04} systems and even change
its sign by doping \cite{Koga02}. In the asymmetric Si/Si$_{1-x}$Ge$_{x}$
quantum wells investigated in Refs.\cite{Jantsch03,Tahan05}, 
where the Dresselhaus term is absent due to the  unit cell inversion symmetry and the 
the band gap is relatively large, the doping-induced  SO  coupling
is three orders of magnitude weaker than in zincblende systems \cite{Sarma05}.

The SO coupling not only provides an ability to manipulate spins with an
electric field, and thus, hopefully, to design a spin transistor \cite{Datta},
but also leads to spin relaxation. The Dyakonov-Perel'
mechanism\cite{Dyakonov72}, which requires random scattering of electrons by impurities
and/or phonons, describes the spin relaxation of non-confined electrons in
the bulk and in two-dimensional (2D) structures where the electron momentum $%
k$ is a relatively well defined quantum number. Here the orientation of the
spin precession axis in the SO field $\hat{H}_{\mathrm{R}}+\hat{H}_{\mathrm{D}}$ changes 
randomly through scattering events. The spin ($\gamma_{\mathrm{DP}}$) and momentum relaxation rates are inversely proportional to each
other in this case. The resulting spin relaxation rate is 
$\gamma_{\mathrm{DP}}\sim\alpha^{2}k^{2}\tau$, with $\tau$ being the momentum relaxation time,
and  $\alpha$ depends on $\alpha_{\mathrm{D}}^{(001)}$ and $\alpha_{\mathrm{R}}$. The spin
relaxation can also arise from the random paths of electrons in regular systems
of antidots \cite{Pershin04} and the tunneling of holes in arrays of Si/Ge QDs 
\cite{Zinovieva05}. Various mechanisms of SO coupling
and spin relaxation were reviewed recently in Ref.\onlinecite{Zutic04}.

For this reason it is necessary to understand the limits to the abilities to
manipulate the electron spins in QDs, where, due to the localization, the
electron momentum is not a good quantum number, and thus the spin dynamics \cite
{Valin02}, and hence the spin relaxation require a more sophisticated
consideration. A possible "admixing" mechanism of spin relaxation in QDs
arises since SO coupling mixes the electron states with opposite spins, and,
therefore, can cause spin-flip transitions by the emission a phonon 
\cite{Khaetskii00,Wu04,Florescu04,Bastard92}. This effect is a manifestation
of an effective spin-phonon coupling caused by the SO interaction. 
It has been shown, however, that other
mechanisms can be important for the spin relaxation. For example, the
interaction of the electron spin with the spins of nuclei in GaAs QDs can
lead to spin relaxation even if no SO coupling in the Rashba or Dresselhaus
form is present \cite{Nuclear}. This interaction limits the maximal spin
relaxation time. In Si-based QDs the spin relaxation can arise due to the
modulation of the electron $g-$factor by phonons \cite{Kim03}. However,
theoretically considered \cite{Sherman03a,Sherman03b,Golub04} and recently 
observed \cite{Jantsch03,Tahan05} Rashba- and Dresselhaus-type SO
coupling in 2D Si systems can lead to a more conventional ''admixing''
mechanism of the spin relaxation there. We note here that the randomness in
SO coupling leads to a Gaussian rather than to an exponential decay of the
spin polarization \cite{Glazov05} as well as to a serious limitation of the
operational modes of the proposed spin transistor devices \cite{Sherman05}.

Thus, the possibility to manipulate the spin states depends on the SO
coupling strength, which, therefore, plays both a positive and negative role
in the spin dynamics of QDs. A solution to this dilemma can be found in the
ability to change the strength of the SO coupling by switching it on and off
when the desired spin states are produced and preserved, respectively. For
example, the effects of the structural asymmetry of a quantum well can be
strongly reduced \cite{Nitta97,Koga02} and/or the Rashba and the Dresselhaus 
terms can be effectively compensated by applying an external bias \cite{Loss02}.
If realized experimentally, the suggestion of Ref.\cite{Loss02},
is expected to open a route for a larger (up to 100 $\mu$m) size spintronic devices.
\cite{Sherman05} 

Here we concentrate on the effect of random SO coupling on spin relaxation
in QDs arising in systems where the artificial asymmetry and, in turn, the
Rashba-type SO coupling is produced by one-sided doping. The randomness of
the SO coupling in our model arises due to fluctuations in the dopant concentration.
Because of the randomness, the SO coupling cannot be completely compensated
by applying an external bias since the bias is uniform as a function of the
in-plane coordinate. We show that even when the bias removes the regular
part of the SO coupling, the spin relaxation rate is still finite due to the
residual random SO coupling. This effect limits a possible decrease of
the spin relaxation rate to two orders of magnitude at most. As another
example of the role of inhomogeneous SO coupling in QDs we note its
influence on the localization effects \cite{Falko03} and on
spin dynamics of electrons in quantum rings interacting with the
surrounding nuclei \cite{Vagner98}.

This paper is organized as follows. Section II presents a model of the
random SO coupling in two-dimensional structures and provides us with the
basis for the calculations for QDs. In the third Section we present a theory of
spin relaxation in QDs with random SO coupling, emphasizing
InGaAs-like structures. A summary of the results and a discussion of possible
extensions of this work is given in the conclusions.  

\section{Uniform and random SO coupling}

In 2D systems where SO coupling arises due to one-sided doping, the 
$\alpha_{\mathrm{R}}$ parameter cannot be considered as a constant in space since
fluctuations of the concentration of the dopant cause its randomness. For
doped Si and Ge bulk crystals the importance of randomness of the SO
coupling was first understood by Mel'nikov and Rashba \cite{Melnikov}. Below
we consider the dopant fluctuations-induced Rashba field in lateral QDs,
calculate the corresponding spin relaxation rate, and show that  for applications 
the randomness imposes important restrictions on the  minimum spin-flip rate.

A structure consisting of a 2D channel where the QDs are formed and a narrow
dopant layer with a 2D concentration $n(\mathbf{r})$ of dopants
with charge $\left| e\right|$ is considered, with $\mathbf{r=(}r_{x},r_{y}%
\mathbf{)}$ being the 2D in-plane radius-vector. As the factor that
determines the local strength of the SO coupling, we consider the $%
z- $component of the electric field of the dopant ions ${E}_{z}({\bm \rho })$
at a point with 2D coordinate ${\bm \rho }$ in the well symmetry plane. We
assume that the SO coupling is a linear function of ${E}_{z}(\mathbf{\rho })$
with $\alpha _{\mathrm{R}}({\bm\rho})=\alpha_{\rm SO}|e|{E}_{z}({\bm \rho })$, where 
$\alpha_{\rm SO}$ is a system-dependent parameter that includes the influence of the
electric field on the polarization of the electron wavefunction in a quantum
well \cite{Silva97,Grundler99}. The $z$-component of the Coulomb field of the dopant
ions is given by 
\begin{equation}
{E}_{z}({\bm \rho })=\frac{\left| {e}\right| }{\epsilon }\int n(\mathbf{r})f(%
{\bm \rho },\mathbf{r})d^{2}r,
\end{equation}
where $\epsilon $ is the dielectric constant. The integration is performed
over the dopant layer. Function $f({\bm \rho },\mathbf{r})$ has the form: 
\begin{equation}
f({\bm \rho },\mathbf{r})=\frac{z_{0}}{[({\bm\rho}-\mathbf{r}%
)^{2}+z_{0}^{2}]^{3/2}}.
\end{equation}
Here ${\bm\rho}=(\rho_{x},\rho_{y})$ is two-dimensional vector
characterizing the electron positions in the conducting layer. We assume
that the correlation function of the dopant concentration is  ``white
noise'' in the form: 
\begin{equation}
\langle (n(\mathbf{r}_{1})-\overline{n})(n(\mathbf{r}_{2})-\overline{n}%
)\rangle =\overline{n}\delta (\mathbf{r}_{1}-\mathbf{r}_{2}).
\end{equation}
Here $\overline{n}=\langle n(\mathbf{r})\rangle $, and $\left\langle
...\right\rangle $ stands for the average. The fluctuations are taken to
obey Gaussian statistics, as is commonly considered in the theory of
doped semiconductors \cite{Efros89}. With the increase of the distance
between the layers, the fluctuations of $\alpha _{\mathrm{R}}({\bm \rho }) $
become smaller and smoother. The total asymmetry-induced Rashba parameter is
a sum of a regular average $\langle \alpha \rangle _{a}$ and a random term
comprising the zero mean ${\bm \rho }$-independent contribution: 
\begin{equation}  \label{eq::alpha}
\alpha _{\mathrm{R}}({\bm \rho})=\langle \alpha \rangle _{0}+\alpha _{\mathrm{rnd}}({\bm \rho }),
\end{equation}
with $\langle\alpha\rangle_{0}=2\pi\alpha_{\rm SO}\overline{n}e^2/\epsilon.$
For the $\delta (z-z_{0})-$doping considered here the
magnitude of the random term is given by 
$\sqrt{\left\langle\alpha _{\mathrm{rnd}}^{2}\right\rangle }=
\alpha_{\rm SO}e^2\sqrt{\pi\overline{n}}/\epsilon z_{0}$
\cite{Sherman03b}. By comparing this result with 
$\langle\alpha\rangle_{0}$ one concludes that randomness of the SO coupling
becomes important when $\sqrt{\langle \alpha _{\mathrm{rnd}}^{2}\rangle }%
\sim \langle \alpha \rangle _{0},$ that is at $\overline{n}z_{0}^{2}\sim
0.1, $ which is a typical parameter value for zincblende quantum wells \cite
{Ando,Winkler}. We assume that by applying an external bias $V$ across the QD
plane the mean asymmetry-induced term $\langle \alpha \rangle _{a}$ is
decreasing as $\langle \alpha \rangle _{a}=\langle \alpha \rangle _{0}\left(
V_{c}-V\right) /V_{c}$ and that a critical bias $V_{c}$, which is of the order
of few Volts \cite{Nitta97,Koga02}, fully compensates the
mean SO coupling.

The Hamiltonian of the coordinate-dependent SO coupling has to be
written in the symmetrized Hermitian form:

\begin{equation}  \label{eq::Hamiltonian}
\hat{H}_{\mathrm{R}}=\frac{1}{2}\left[ \sigma _{x}\left\{ \hat{k}_{y},\alpha
_{\mathrm{R}}\left( {\bm \rho }\right) \right\} -\sigma _{y}\left\{ \hat{k}%
_{x},\alpha _{\mathrm{R}}\left( {\bm \rho }\right) \right\} \right] ,
\end{equation}
where $\left\{ \hat{k}_{i},\alpha _{\mathrm{R}}\left( {\bm \rho }\right)
\right\} $ stands for the anticommutator.

Quantitatively, the spatial behavior of $\alpha _{\mathrm{R}}({\bm \rho }) $
is characterized by the correlation function $F_{\mathrm{\alpha\alpha}}\left(\tilde{{\bm\rho}}_{12}\right),$ where 
$\tilde{\bm\rho}_{12}\equiv \left( {\bm \rho }_{1}-{\bm\rho}_{2}\right) /z_{0},$ of
the random Rashba parameter determined by: 
\begin{equation}  \label{eq::correlatorFaa}
\langle \alpha _{\mathrm{R}}({\bm \rho }_{1})
\alpha_{\mathrm{R}}({\bm\rho}_{2}))\rangle =\left\langle \alpha _{\mathrm{rnd}}^{2}\right\rangle F_{%
\mathrm{\alpha \alpha }}\left( \tilde{{\bm \rho }}_{12}\right).
\end{equation}
For the Gaussian fluctuations (see, for example Refs. \cite
{Sherman03a,Efros89}), the correlation function in Eq.(\ref
{eq::correlatorFaa}) necessary for describing spin relaxation of
semiclassically moving itinerant electrons \cite{Sherman03a} is: 
\begin{equation}
F_{\mathrm{\alpha \alpha }}\left( \tilde{{\bm \rho }}_{12}\right) =\frac{%
2}{\pi }\int \frac{1}{\left[ r^{2}+1\right] ^{3/2}\left[ \left( \mathbf{r-}%
\tilde{{\bm \rho }}_{12}\right) ^{2}+1\right] ^{3/2}}d^{2}r.
\end{equation}

For the following investigations of the spatially random SO coupling in
quantum systems \cite{Mireles02} described by the Hamiltonian in Eq.(\ref
{eq::Hamiltonian}) we need additional correlation functions of the random
Rashba field and its derivatives, namely ($i,j$ are the Cartesian indices $%
x,y$):

\begin{eqnarray}
\left\langle \alpha _{\mathrm{R}}({\bm \rho }_{1})\frac{\partial \alpha _{%
\mathrm{R}}({\bm \rho }_{2})}{\partial \rho _{2}^{(i)}}\right\rangle &\equiv
&\left\langle \alpha _{\mathrm{rnd}}^{2}\right\rangle \frac{F_{\alpha \alpha
_{i}}\left( \tilde{{\bm \rho }}_{12}\right) }{z_{0}},\qquad  \nonumber \\
\left\langle \frac{\partial \alpha _{\mathrm{R}}({\bm \rho }_{1})}{\partial
\rho _{1}^{(i)}}\alpha _{\mathrm{R}}({\bm \rho }_{2})\right\rangle &\equiv
&\left\langle \alpha _{\mathrm{rnd}}^{2}\right\rangle \frac{F_{\alpha
_{i}\alpha }\left( \tilde{{\bm\rho}}_{12}\right) }{z_{0}},  \nonumber \\
\left\langle \frac{\partial \alpha _{\mathrm{R}}({\bm \rho }_{1})}{\partial
\rho _{1}^{(i)}}\frac{\partial \alpha _{\mathrm{R}}({\bm \rho }_{2})}{%
\partial \rho _{2}^{(j)}}\right\rangle &\equiv &\left\langle \alpha _{%
\mathrm{rnd}}^{2}\right\rangle \frac{F_{\alpha _{i}\alpha _{j}}\left( 
\tilde{{\bm \rho }}_{12}\right) }{z_{0}^{2}}.\qquad
\end{eqnarray}
The dimensionless correlators can be calculated, for example, as follows:

\begin{eqnarray}
F_{\alpha \alpha _{x}}\left( \tilde{{\bm \rho }}_{12}\right) &=&\frac{6}{%
\pi }\int \frac{r_{x}}{\left[ r^{2}+1\right] ^{5/2}\left[ \left( \mathbf{r-}%
\tilde{{\bm \rho }}_{12}\right) ^{2}+1\right] ^{3/2}}d^{2}r,\qquad \\
&&  \nonumber \\
F_{\alpha _{y}\alpha _{x}}\left( \tilde{{\bm \rho }}_{12}\right) &=&%
\frac{18}{\pi }\int \frac{r_{y}\left( \mathbf{r-}\tilde{{\bm \rho }}%
_{12}\right) _{x}}{\left[ r^{2}+1\right] ^{5/2}\left[ \left( \mathbf{r-}%
\tilde{{\bm \rho }}_{12}\right) ^{2}+1\right] ^{5/2}}d^{2}r.  \nonumber
\end{eqnarray}
Other components of the correlation functions can be obtained from 
$x\longleftrightarrow y$ permutations. We mention two important properties of
the correlators concerning their behavior at small and large distances. The
first property is that $F_{\alpha \alpha _{i}}-$type correlators vanish at ${\bm\rho}_{1}={\bm\rho}_{2}.$
The second property is the fast decay of the long-distance asymptotic
of the correlators given, for example, by: 
\begin{eqnarray}
F_{\alpha\alpha }\left({\rho}_{12}\gg 1\right)        &\sim &\frac{1}{\tilde{\rho}_{12}^{3}},  \nonumber \\
F_{\alpha\alpha _{x}}\left( {\rho}_{12}\gg 1\right)   &\sim &\frac{n_{12}^{x}}{\tilde{\rho}_{12}^{4}},  \nonumber \\
F_{\alpha_{x}\alpha_{y}}\left({\rho}_{12}\gg 1\right) &\sim &\frac{n_{12}^{x}n_{12}^{y}}{\tilde{\rho}_{12}^{5}},
\end{eqnarray}
with the unit vector $\mathbf{n}_{12}=\tilde{{\bm\rho}}_{12}/\tilde{\rho}_{12}.$ The
correlation functions shown in Fig.1 decay at $\rho_{12}\sim 1,$ and
therefore, establish a general spatial nanoscale for the lateral and the $z-$%
axis directions.

In zincblende-based structures the Dresselhaus SO term arising from
the unit cell inversion asymmetry has to be added to the Rashba term. In the 
(001) structures the
coupling parameter $\alpha_{\mathrm{D}}^{(001)}=\alpha_{c}(\pi/w)^{2},$ where 
$\alpha_{c}$ is the bulk Dresselhaus coupling constant ($\approx 25$ eVA$^{3}
$ in GaAs and InAs [\onlinecite{Winkler}]), and $w$ is the quantum well width. We consider below a
model where the random contribution to the Rashba coupling $\left\langle
\alpha_{\mathrm{rnd}}^{2}\right\rangle $ is much larger than 
$\left(\alpha_{\mathrm{D}}^{(001)}\right)^{2},$ which is possible for sufficiently broad quantum wells,
and, therefore, neglect the $\hat{H}_{\mathrm{D}}^{(001)}$ term in the calculation 
procedure and subsequently discuss the role of this term in relation to our results.

Another possibility is provided by (011) QWs, where the Dresselhaus term \cite{Dyakonov86}
has the form: 
\begin{equation}
{H}_{D}^{(011)}=\alpha_{D}^{(011)}k_{y}{\sigma }_{z}
\left[1-\left(k_{y}^{2}-2k_{x}^{2}\right)w^2\pi^2\right].
\end{equation}
Here the $z-$axis is perpendicular to the QW plane and the in-plane axes are: $%
x=[100]$ and $y=[0\overline{1}1]$, and $\alpha_{D}^{(011)}=\alpha_{c}(\pi/w)^2/2$. 
The $H_D^{(011)}$ coupling proportional
to $\sigma_{z}$ does not lead to the electron spin-flip, and, therefore, 
the main contribution to the
spin relaxation rate comes from the randomness.  

The strength of the Dresselhaus 
coupling can be considerably decreased by the strain induced in QWs by a mismatch
in the parental lattices, which, therefore, can decrease the contribution
of the Dresselhaus terms in the spin relaxation rate in (001) 
structures.\cite{Wu05}

\section{Longitudinal spin relaxation rate}

First, we briefly review the influence of the random SO coupling on the spin
relaxation of non-confined electrons in quantum wells. The spin precession
rate and direction at a point ${\bm \rho }$ are determined by the random
local coupling $\alpha _{\mathrm{R}}\left( {\bm \rho }\right) $. As a
result, the spin precession is random even for a carrier moving
straightforwardly and the total spin of the electron ensemble relaxes even if
the regular term $\langle \alpha \rangle _{a}$ vanishes as a result of
applying either an external bias or equivalent doping at the other side of
the conducting layer \cite{Sherman03a}. 

Now we consider a single-electron QD produced by applying a lateral bias
to a 2D electron gas (see Fig. 2) leading to electron localization. From the experimental point of
view, we shall concentrate on In$_{0.5}$Ga$_{0.5}$As$-$like systems, where
the ability to manipulate the SO coupling by a moderate bias applied
across the structure has been clearly proven \cite{Nitta97,Koga02}. 

The spectrum of an electron in the QD, $E_{n,\ell _{z}},$ is determined both
by the confinement potential $m^{*}\rho ^{2}\omega ^{2}/2$ (Fig. 2 shows
two quantum dots on the random Rashba SO coupling background)
and the uniform static magnetic field $\mathbf{B}\parallel z$ with the axial
gauge $\mathbf{A}=[\mathbf{B},\mathbf{r]}/2$, respectively, leading to 
\begin{equation}
E_{n,\ell _{z}}=-\ell _{z}\frac{\hbar^{2}}{2m^{*}l_{B}^{2}}+\frac{\hbar ^{2}%
}{2m^{*}a^{2}}\left(1+\left|\ell_{z}\right| +2n\right) \pm \frac{g\mu _{B}%
}{2}B,  \label{eq::spectrum}
\end{equation}
where $m^{*}$ is the lateral electron effective mass, 
$\ell_{z}$ is the $z-$component of the orbital momentum, $n$ is the radial quantum number, and the
corresponding wave function

\begin{equation}
\psi _{n\ell _{z}}({\bm \rho })\left| s\right\rangle =\frac{1}{\sqrt{2\pi }a}%
\exp (-\rho ^{2}/4a^{2})L_{n\ell _{z}}(\rho /a)e^{i\ell _{z}\varphi }\left|
s\right\rangle .
\end{equation}
Here $\varphi $ is the azimuthal angle, the magnetic length $l_{B}= \sqrt{%
\hbar c/\left| e\right| B}$, $L_{n\ell_{z}}(\rho /a)$ is a Laguerre
polynomial, and the spatial scale $a=l_{0}l_{B}\left(
l_{0}^{4}+l_{B}^{4}\right) ^{-1/4},$ with $l_{0}=\sqrt{\hbar /2m^{*}\omega }$
being the ''oscillator'' length in the absence of a magnetic field due to
the lateral confinement forming the QD, $\hbar \omega $ is the separation of
the energy levels at $B=0$, and $\left| s\right\rangle =\left| \uparrow
\right\rangle $,$\left| \downarrow \right\rangle $ is the spin state. With
the increase of magnetic field, $E_{n,\ell_{z}}$ changes from the spectrum
of a single-electron QD to the spectrum of a free electron in a magnetic
field.

We consider below the QD in only the orbital ground state with the function 
\begin{equation}
\psi _{00}({\bm \rho })=\frac{1}{\sqrt{2\pi }a}\exp (-\rho
^{2}/4a^{2}).\qquad 
\end{equation}
The spin-flip-transitions such as 
$\psi_{00}({\bm\rho})\left|\uparrow\right\rangle\rightarrow\psi_{00}({\bm\rho})\left|\downarrow\right\rangle$
without change of the orbital quantum numbers occur at a
frequency $\omega=\left|g\right|\mu _{B}B/\hbar $ and the transition energy is
released to an acoustic phonon with momentum $q=\left|g\right|\mu_{B}B/c_{\lambda}\hbar,$ 
where $c_{\lambda}$ is the sound velocity for the
given phonon branch $\lambda$ with $\lambda=\parallel$ and $\lambda=\perp$ for the longitudinal
and the transverse modes, respectively. The electron-phonon coupling Hamiltonian for
acoustic phonons $\hat{H}_{\mathrm{e-ph}}$ has the form
\[
\hat{H}_{\mathrm{e-ph}}=\sum_{\lambda }\left( \hat{H}_{\mathrm{e-ph,}%
P}^{[\lambda ]}+\hat{H}_{\mathrm{e-ph,}D}^{[\lambda ]}\right) ,
\]
arising due to the piezoeffect  
$\left(\hat{H}_{\mathrm{e-ph,}P}^{[\lambda]}\right)$ and due to the deformation potential  
$\left(\hat{H}_{\mathrm{e-ph,}D}\right).$ These terms have the form (assuming
the crystal volume equal to one): 
\begin{widetext}
\begin{eqnarray}
\hat{H}_{\mathrm{e-ph},P}^{[\lambda ]} &=&\sqrt{2\pi }\frac{\kappa _{14}}{%
\sqrt{\zeta }}\frac{e}{\epsilon }\sum_{\mathbf{q,e}}\frac{1}{\sqrt{%
qc_{\lambda }}}\left[ e_{x}e_{y}d^{z}+e_{z}e_{x}d^{y}+e_{y}e_{z}d^{x}\right]
\left( e^{-i\mathbf{qr}}\hat{b}_{\lambda ,\mathbf{q,e}}^{\dagger }+%
\mathrm{h.c}\right) , \nonumber \\
\hat{H}_{\mathrm{e-ph,}D}^{[\lambda ]} &=&\frac{D\sqrt{\hbar }}{\sqrt{2\zeta
c_{\lambda }}}\sum_{\mathbf{q,e}}\sqrt{q}\left( \mathbf{de}\right) \left(
e^{-i\mathbf{qr}}\hat{b}_{\lambda ,\mathbf{q,e}}^{\dagger }+\mathrm{h.c}%
\right) ,
\end{eqnarray}
\end{widetext}
where the summation is taken over the phonon momenta $\mathbf{q}$,
and polarization $\mathbf{e}.$ The propagation direction $\mathbf{d=q}/q$, $\kappa
_{14}$ is the strength of the piezointeraction, $D$ is the deformation
potential, $\zeta $ is the crystal
density, $\epsilon $ is the dielectric constant, and $\hat{b}_{\lambda ,%
\mathbf{q,e}}^{\dagger }$ is the phonon creation operator. $\hat{H}_{\mathrm{e-ph,}D}^{[\lambda ]}$ 
is nonzero for the longitudinal phonon mode ($\mathbf{de}=1$)
only. At finite temperature, the application of Fermi's golden
rule yields the mean value of the spin relaxation rate $\left\langle \Gamma
_{1}\right\rangle \equiv \left\langle T_{1}^{-1}\right\rangle $, where $T_{1}
$ is the relaxation time, with respect to the spin-flip as: 
\begin{widetext}
\begin{equation}
\left\langle \Gamma _{1}\right\rangle =\frac{2\pi }{\hbar }\left[
n_{B}+1\right] \sum_{\lambda }\nu _{\mathrm{ph}}^{[\lambda ]}\left( \left|
g\right| \mu _{B}B\right) \left\langle \left| \left\langle \tilde{s}%
^{\prime }\right| \tilde{\psi }_{\mathrm{gr}}({\bm \rho })\left| \hat{H}%
_{\mathrm{e-ph}}^{[\lambda ]}\right| \tilde{\psi }_{\mathrm{gr}}(\mathbf{%
\rho })\left| \tilde{s}\right\rangle \right| ^{2}\right\rangle _{\mathbf{%
d,e}},
\end{equation}
\end{widetext}
where $n_{B}=$ $\left[ \exp \left( \left| g\right| \mu _{B}B/T\right)
-1\right] ^{-1}$ is the Bose-Einstein phonon occupation factor. We consider
below only the zero temperature case for simplicity. The phonon density of states 
$\nu _{\mathrm{ph}}^{[\lambda ]}\left( \varepsilon \right) $ $=\varepsilon
^{2}/2\pi ^{2}(\hbar c_{\lambda })^{3},$ and $\left\langle ...\right\rangle
_{\mathbf{d,e}}$ stands for the average over the phonon directions and
polarizations. The spin-flip matrix element of $\hat{H}_{\mathrm{e-ph}}$ is
nonzero due to the admixing of the upper $\psi _{n\ell _{z}}({\bm \rho }%
)\left| s\right\rangle $ states to the ground state $\tilde{\psi }_{%
\mathrm{gr}}{\bm\rho})\left| \tilde{s}\right\rangle,$ where $%
\left| \tilde{s}\right\rangle $ shows that spin is an approximate
quantum number due to the SO coupling. Here we have neglected the spin
splitting of the states $g\mu_{B}B$ in comparison to the energy corresponding to the 
orbital degrees of freedom, 
$\hbar\omega_{n\ell_{z}}$. The approximation is a reasonable one 
despite the large $g-$factor in In$_{0.5}$Ga$_{0.5}$As ($g\approx 4.0$)
quantum wells,  since a small effective mass $\left( m^{*}=0.04m_{0}\right)$
leads to $g\mu_{B}B/\hbar\omega_{n\ell_{z}}\le 0.2$. 
We mention here that due to the random
coordinate dependence of the Hamiltonian $\hat{H}_{\mathrm{R}}$, there are no
symmetry-related selection rules for $n$ and $\ell_{z}$ in the matrix
elements 
$\left\langle s^{\prime}\right|\psi_{n\ell_{z}}({\bm\rho})
\left|\hat{H}_{\mathrm{R}}\right|\psi_{00}({\bm\rho})\left|s\right\rangle$ that are
responsible for the formation of the spin relaxation rate, in contrast to the case of
regular SO coupling \cite{Wu04}. Since the Hamiltonians $\hat{H}_{\mathrm{%
e-ph}}^{[\lambda ]}$ and $\hat{H}_{\mathrm{e-ph,}D}^{[\lambda ]}$ have
drastically different phonon momentum dependencies, the relative
contribution of the deformation potential mechanism increases with an
increase in the magnetic field \cite{Alcalde04}. Due to the large $g$-factor, this mechanism
already dominates at $B=0.1$ T.

By using the approach suggested in Ref.\cite{Khaetskii00} and taking into
account the Gaussian character of the fluctuations, after some algebraic
transformations we obtain for the deformation potential contribution
using the notation  
$\mathbf{0}\equiv \left(0,0\right),\mathbf{1}\equiv \left( n_{1},\ell _{1}^{z}\right) ,\mathbf{2}%
\equiv \left( n_{2},\ell _{2}^{z}\right):$
\begin{widetext}
\begin{eqnarray}
\left\langle \Gamma _{1}\right\rangle  &=& 
\frac{\left(g\mu_{B}\right)^{5}}{4\hbar^{4}} 
\frac{D^{2}}{\zeta c_{\parallel}^{5}}B^{5}
\sum_{\mathbf{1,2}}\frac{1}{\omega _{n_{1},\ell
_{1}^{z}}^{2}\omega _{n_{2},\ell _{2}^{z}}^{2}}\times  \\
&&\hspace{-4cm}\left[ \left\langle K_{-}(\mathbf{1})K_{-}^{*}(\mathbf{2}%
)\right\rangle \left\langle g_{\parallel }(\mathbf{0};\mathbf{1}%
)g_{\parallel }^{*}(\mathbf{0};\mathbf{2})\right\rangle _{\mathbf{d}%
}+\left\langle K_{-}(\mathbf{1})K_{+}^{*}(\mathbf{2})\right\rangle
\left\langle g_{\parallel }(\mathbf{0};\mathbf{1})g_{\parallel }^{*}(\mathbf{%
2};\mathbf{0})\right\rangle _{\mathbf{d}}\right. +  \nonumber \\
&&\hspace{-4cm}\left. \left\langle K_{+}(\mathbf{1})K_{-}^{*}(\mathbf{2}%
)\right\rangle \left\langle g_{\parallel }(\mathbf{1};\mathbf{0}%
)g_{\parallel }^{*}(\mathbf{0};\mathbf{2})\right\rangle _{\mathbf{d}%
}+\left\langle K_{+}(\mathbf{1})K_{+}^{*}(\mathbf{2})\right\rangle
\left\langle g_{\parallel }(\mathbf{1};\mathbf{0})g_{\parallel }^{*}(\mathbf{%
2};\mathbf{0})\right\rangle _{\mathbf{d}}\right] .  \nonumber
\end{eqnarray}
\end{widetext}
The correlators are given by : 
\begin{eqnarray*}
\left\langle K_{-}(\mathbf{1})K_{-}^{*}(\mathbf{2})\right\rangle 
&=&K_{yy}+K_{xx}-iK_{xy}+iK_{yx}, \\
\left\langle K_{+}(\mathbf{1})K_{+}^{*}(\mathbf{2})\right\rangle 
&=&K_{yy}+K_{xx}+iK_{xy}-iK_{yx}, \\
\left\langle K_{-}(\mathbf{1})K_{+}^{*}(\mathbf{2})\right\rangle 
&=&K_{yy}-K_{xx}-iK_{xy}-iK_{yx}, \\
\left\langle K_{+}(\mathbf{1})K_{-}^{*}(\mathbf{2})\right\rangle 
&=&K_{yy}-K_{xx}+iK_{xy}+iK_{yx},
\end{eqnarray*}
with the matrix elements averaged over the disorder: 
\begin{widetext}
\begin{eqnarray}
K_{ji} &=&\left\langle \alpha _{\mathrm{rnd}}^{2}\right\rangle \int \psi
_{n_{2}\ell _{2}^{z}}({\bm \rho }_{2})\psi _{n_{1}\ell _{1}^{z}}({\bm\rho}_{1})\times\nonumber\label{eq::Kij} \\
&&\hspace{-4cm}\left\{ F_{\alpha \alpha }({\bm \rho }_{12})\hat{\overline{P}}%
_{j}^{(2)}\hat{P}_{i}^{(1)}+\frac{i}{z_{0}}\left( F_{\alpha _{j}\alpha }({%
\bm \rho }_{12})\hat{P}_{i}^{(1)}-F_{\alpha \alpha _{i}}({\bm \rho }_{12})%
\hat{\overline{P}}_{j}^{(1)}\right) +\frac{F_{\alpha _{j}\alpha _{i}}({\bm %
\rho }_{12})}{z_{0}^{2}}\right\} \cdot   \nonumber \\
&&\psi _{00}({\bm \rho }_{2})\psi _{00}({\bm \rho }_{1})d^{2}\rho
_{1}d^{2}\rho _{2}+  \nonumber \\
&&\left\langle \alpha \right\rangle _{a}^{2}\int \psi _{n_{2}\ell _{2}^{z}}({%
\bm \rho }_{2})\psi _{n_{1}\ell _{1}^{z}}({\bm \rho }_{1})\hat{\overline{P}}%
_{j}^{(2)}\hat{P}_{i}^{(1)}\psi _{00}({\bm \rho }_{2})\psi _{00}({\bm \rho }%
_{1})d^{2}\rho _{1}d^{2}\rho _{2}.
\end{eqnarray}
\end{widetext}
The operators $\hat{P}_{i}^{(1)},\hat{P}_{j}^{(2)}$ in Eq.(\ref{eq::Kij})
are defined according to 
\begin{equation}
P_{x}^{(1)}=-\frac{\rho _{1y}}{l_{B}^{2}}-2i\frac{\partial }{\partial \rho
_{1x}},\qquad P_{y}^{(1)}=\frac{\rho _{1x}}{l_{B}^{2}}-2i\frac{\partial }{%
\partial \rho _{1y}},
\end{equation}
and act on the wavefunction $\psi_{00}({\bm\rho}_{1})$ and $\psi_{00}({\bm\rho}_{2})$, respectively.

The matrix elements $g_{\lambda}(n,\ell _{z};\mathbf{0})$ of the phonon
emission due to electron-phonon coupling are

\begin{equation}
g_{\lambda }(n,\ell _{z};\mathbf{0})=\int \psi _{0}^{2}(z)\psi _{n,\ell
_{z}}^{*}({\bm \rho })e^{i\mathbf{q}_{\lambda }\mathbf{r}}\psi _{00}({\bm %
\rho })d^{3}r.
\end{equation}
where $\psi_{0}(z)$ corresponds to the $z-$axis quantization and is assumed
here for simplicity to be the rigid-walls wave function 
$\psi_{0}(z)=\sqrt{2/w}\sin(\pi z/w)$. Since we consider broad quantum wells, where the size
quantization energy is relatively small and the electron wavefunction weakly penetrates
into the interfaces, the rigid-walls boundary conditions are justified.

We now consider two applications of the model described above. First, we
consider the effect of the fully compensated Rashba coupling ($V=V_{c}$)
such that we are left  only with the random contribution $\alpha _{\mathrm{rnd}}({%
\bm \rho })$ with $\left\langle\alpha_{\mathrm{rnd}}({\bm\rho})\right\rangle=0$. Fig. 3 
presents the results of numerical calculations (including a small contribution 
of the piezoeffect interaction) of the
spin relaxation rate for In$_{0.5}$Ga$_{0.5}$As quantum dots as a function
of the magnetic field $B$ using Eqs.(\ref{eq::spectrum})-(\ref{eq::Kij}) assuming $D=7.5$ eV 
(Ref.[\onlinecite{Dpot}]), $\kappa _{14}=-0.1$ Cm$^{-2}$ (Ref.[\onlinecite{Gantmakher87}]), and sound velocities $c_{\parallel}=4.7\cdot 10^{5}$ cm/s and 
$c_{\perp}=2.8\cdot 10^{5}$ cm/s for the longitudinal and transverse phonons, respectively 
(Ref. [\onlinecite{Gantmakher87}]). As
can be seen, the spin-flip rate increases very rapidly with the applied
field, analogous to the case of regular SO coupling \cite{Khaetskii00,Alcalde04}.

Second, we investigate the bias dependence of the spin relaxation rate 
$\left\langle \Gamma _{1}\left( V\right) \right\rangle $ corresponding to the
linear dependence of the mean Rashba parameter $\langle \alpha \rangle _{a}$
on the applied bias. In this case, it follows from Eq.(\ref{eq::Kij}),
that this dependence can be presented in the form: 

$$
\hspace{-6cm}\frac{\left\langle \Gamma _{1}(0)\right\rangle }{\left\langle \Gamma
_{1}\left( V\right) \right\rangle }=
$$
\begin{equation}
\frac{\left\langle \Gamma
_{1}(0)\right\rangle }{\left\langle \Gamma _{1}\left( V_{c}\right)
\right\rangle +\left[ \left\langle \Gamma _{1}(0)\right\rangle -\left\langle
\Gamma _{1}\left( V_{c}\right) \right\rangle \right] \left( \left(
V_{c}-V\right) /V_{c}\right) ^{2}}.
\end{equation}
Figure 4 shows the $\left\langle\Gamma_{1}(0)\right\rangle /\left\langle
\Gamma _{1}\left( V\right) \right\rangle $ ratio for the two structures
presented in Fig.3. As can be seen in Fig.4, the
resulting $\left\langle \Gamma _{1}(V_{c})\right\rangle $, (when there is
full compensation $\left( \langle \alpha \rangle _{a}=0\right),$) for 
$z_{0}=10$ nm is approximately 60 times smaller than $\left\langle \Gamma
_{1}(0)\right\rangle $. The residual relaxation rate decreases rapidly with
the increase of the dopant-plane layer distance, corresponding to a decrease
in the random fluctuations of the SO coupling.

We note that the randomness of $\alpha _{\mathrm{rnd}}({\bm \rho })$ leads
to an inhomogeneous broadening $\sigma _{\Gamma }=\left\langle \left( \Gamma
_{1}-\left\langle \Gamma _{1}\right\rangle \right) ^{2}\right\rangle ^{1/2}$
of the spin-flip transitions making the relaxation rate dependent on the QD
position. For the Gaussian fluctuations the broadening $\sigma _{\Gamma }$
becomes of the order of the mean value of $\Gamma _{1}$ at full compensation (%
$V=V_{c}$). The effect of inhomogeneous broadening can be seen clearly when
the lateral confinement of the wave function $a$ is considerably smaller than the
spatial scale of the fluctuations of the random potential $z_{0}.$ In this
case each QD in the ensemble interacts mostly with the local Rashba field 
$\alpha({\bm\rho}_{0})$, where ${\bm\rho}_{0}$ is the position of the QD,
that in this case varies weakly on the spatial scale of $a.$ In the case of
full compensation, the relaxation rate for each QD is proportional to
the local $\alpha _{\mathrm{rnd}}^{2}({\bm \rho }_{0})$ leading to an
inhomogeneous width of the transition of the order of $\left\langle \alpha _{%
\mathrm{rnd}}^{2}\right\rangle,$ being, therefore, of the order of $%
\left\langle \Gamma _{1}\right\rangle $ itself. Experimentally, this
situation can be realized, for example, in strong magnetic fields $B>10$ T,
where the magnetic length $l_{B}<10$ nm. In the $a\ll z_{0}$ limit the ratio
of the relaxation rates, which can be found from the discussion following Eq.(\ref{eq::alpha}), 
becomes $\left\langle \Gamma _{1}(V_{c})\right\rangle/\left\langle \Gamma _{1}\left( 0\right) \right\rangle =\left\langle \alpha
_{\mathrm{rnd}}^{2}\right\rangle /\left\langle \alpha \right\rangle
_{0}^{2}=1/8\pi \overline{n}z_{0}^{2}.$

We now discuss the relative role of the random Rashba and Dresselhaus SO
interactions since both effects contribute to the relaxation rate at  $%
V=V_{c}.$ In an In$_{0.5}$Ga$_{0.5}$As quantum dot of the width $w=$15 nm,
as presented in Figs. 3 and 4, the Dresselhaus term $\alpha_{D}^{(001)}$ is of
approximately $1.25\times 10^{-10}$ eVcm. At the same time, the one-side
doping  with $\overline{n}=8\times 10^{11}$ cm$^{-2}$ leads to $\left\langle
\alpha \right\rangle _{0}\approx 10^{-9}$ eVcm, giving
in this case $\left\langle \Gamma _{1}(0)\right\rangle /\left\langle \Gamma
_{1}\left( V_c\right) \right\rangle \sim 60.$ Therefore, the Dresselhaus and
the random Rashba terms give similar contributions to the spin relaxation
under these conditions. The ratio of these contributions depends on the 
applied magnetic field, and numerical analysis shows that the random part can
dominate over the regular one. For  (011) structures the random part gives
the major contribution. 

\section{Conclusions}

In this analysis we have found that due to fluctuations of the SO coupling,
the spin relaxation in QDs does not vanish even in the case of full
compensation of the mean value of the SO coupling by an external bias. The
residual spin relaxation rate is of the order of few percent of the spin
relaxation rate in the case of a non-compensated field and strongly depends
on the system properties, the size of the dot, and the applied magnetic
field. In the case of full compensation, the inhomogeneous line width
of the spin-flip transition is of the order of the mean line frequency, and,
therefore, one has a broad distribution of the relaxation rates over the
ensemble of QDs. The effect of the random Rashba SO coupling is comparable
to or larger than the effect of the Dresselhaus coupling. The ratio of the
contributions of these SO coupling mechanisms to the spin relaxation rate is compound- and
structure-dependent, and shows that the importance of the randomness of the
Rashba SO coupling. 

We concentrated here on the longitudinal spin relaxation time 
$T_{1}=\Gamma_{1}^{-1}$ that is related to the energy release to the acoustic
phonon bath in the spin-flip process. However, the spin ensemble dephasing
time $T_{2},$ describing the motion of the spin component perpendicular to
the magnetic field, and which may not be related to the energy transfer
from the spin system to the lattice, is an important characteristic of a QD
ensemble. As was recently found theoretically, the physics behind these
two relaxation processes in systems of QDs coupled to phonons is similar \cite
{Golovach04}. Therefore, the randomness of the SO coupling considered above also
influences the dephasing time $T_{2}$. This is an interesting  problem
which will be investigated separately. The role of the magnetic field parallel to the plane \cite{Falko05}
of the QD  would be another extension of the results  presented here. 

In this paper we have discussed zincblende quantum dots only. However, some
remarks concerning Si-based structures need to be given. If the admixing
mechanism of spin relaxation in these structures becomes important, the
arguments discussed  above concerning the role of the
randomness could be applied there to an even larger extent due to the absence of
the Dresselhaus term. A quantitative analysis of this problem necessitates a 
comparison of the effectiveness of different
mechanisms of spin relaxation in Si-based QDs, which is a separate
interesting issue.

\section{Acknowledgment}

This work was supported by the DARPA SpinS program and the Austrian Science
Fund FWF via grant P15520. EYS is grateful to J.E. Sipe for very valuable
discussions.

\newpage


\newpage

\begin{widetext}

\begin{figure}[tbp]
\includegraphics[width=8.0cm]{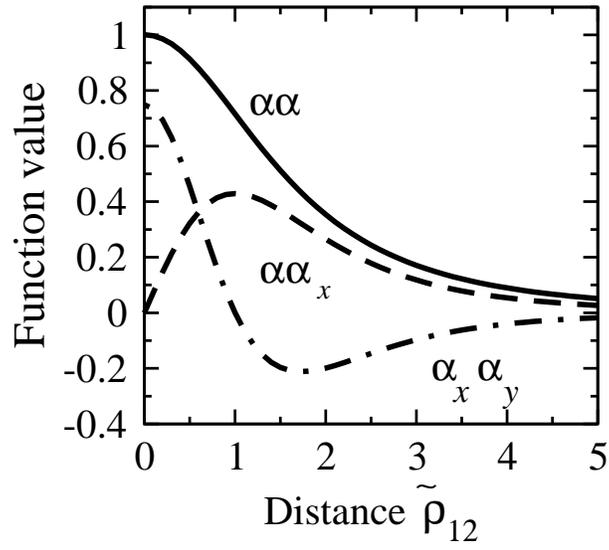} \vspace{10cm}
\caption{ Correlation functions of the random SO coupling parameter $\alpha
_{\mathrm{R}}({\bm \rho })$ and its derivatives: $F_{\alpha\alpha}$ $(%
\tilde{{\bm\rho }}_{12})$ (solid line), $F_{\alpha \alpha _{x}}$ $(%
\tilde{{\bm\rho }}_{12})$ (dashed line), $F_{\alpha _{x}\alpha _{y}}$ $(%
\tilde{{\bm\rho }}_{12})$ (dash-dotted line); $\tilde{\bm\rho}_{12}$
is parallel to the $[110]$ axis.}
\end{figure}

\newpage

\begin{figure}[tbp]
\includegraphics[width=8.0cm]{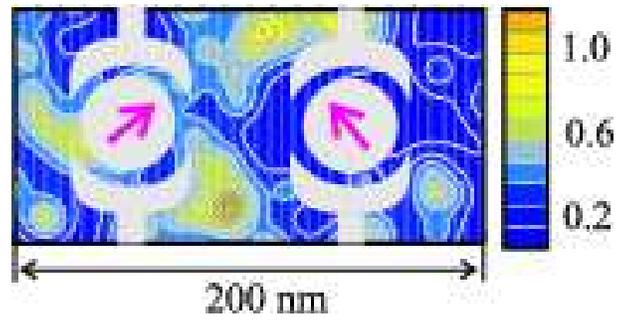}
\caption{(Color online) Two quantum dots on a template with a random Rashba SO
coupling. The arrows show schematically the spin orientations. The gates 
form the confining in-plane potential. The scale on the right represents the
Rashba parameter in 10$^{-9}$ eVcm units, $\langle\alpha\rangle=0.5\times
10^{-9}$ eVcm, $\overline{n}=2.5\times 10^{11}$ cm$^{-2}$, and $z_0=10$ nm.}
\end{figure}

\newpage

\begin{figure}[tbp]
\vspace{10cm} \includegraphics[width=8.0cm]{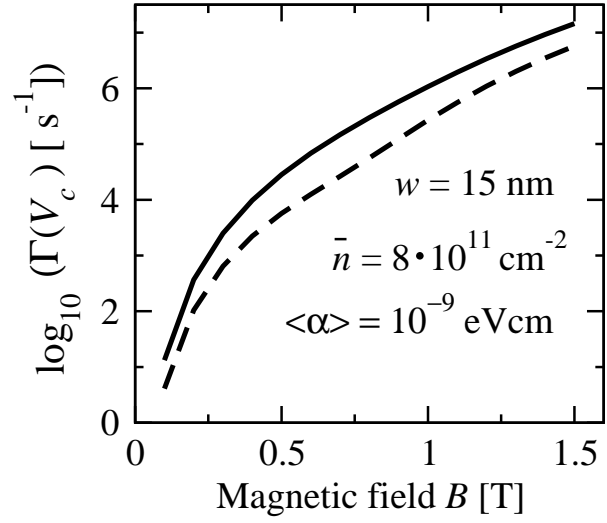}
\caption{Logarithm of the spin relaxation rate as a function of applied magnetic field
for different $z_{0}$ in the case of total compensation of the regular SO
coupling by an external bias ($V=V_{c}$); $z_{0}=10$ nm (solid line), $z_{0}=20$
nm (dashed line), and $l_{0}=15$ nm. The relationship between the doping 
$\overline{n}$ and Rashba constant $\langle\alpha\rangle$ corresponds to the
experimental data of Ref. [\onlinecite{Koga02}].}
\end{figure}

\newpage

\begin{figure}[tbp]
\includegraphics[width=8.0cm]{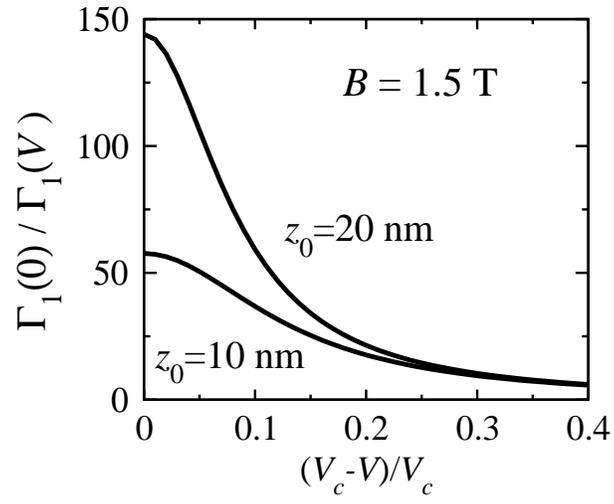}
\caption{ The ratio of longitudinal spin relaxation rates $\left\langle
\Gamma_{1}(0)\right\rangle/\left\langle\Gamma_{1}(V)\right\rangle$ as a
function of the applied bias that is gradually compensating the SO coupling. The applied
magnetic field $B=1.5$ T.}
\end{figure}

\end{widetext}

\end{document}